\documentstyle[eqsecnum,preprint,aps]{revtex}

\def\footnoterule{\kern-3pt \hrule width \hsize \kern6.2pt}

\begin{document}

\baselineskip=14pt

\title{Time Reversal Symmetry Breaking Effects\\
in Resonant Nuclear Reactions\footnote[1]{\baselineskip=14pt
This work is
supported in part by funds provided by the U.S.~Department of Energy (D.O.E.)
under cooperative agreement DE-FC02-94ER40818}
}
\author{H.~Feshbach,
M.S.~Hussein\footnote[2]{\baselineskip=14pt
Permanent address:  Instituto de F\'\i sica,
Universidade de S\~ao Paulo, C.P.~20516, S\~ao Paulo, S.P.~Brazil.  Supported
partly by FAPESP (Brazil)},
A.K.~Kerman
}

\medskip

\address{
Center for Theoretical Physics\\
\baselineskip=17pt
Laboratory for Nuclear Science and\\
Department of Physics\\
Massachusetts Institute of Technology\\
Cambridge, MA\ \ 02139\ \ USA
\vskip 3ex}
\date{CTP\#2310 \qquad   October, 1994}

\maketitle

\setcounter{page}{0}
\thispagestyle{empty}

\bigskip \bigskip \bigskip

\centerline{To be published in: {\it Z. Phys.~A\/}}

\bigskip
\begin{abstract}
\baselineskip=14pt

We incorporate time reversal symmetry breaking (TRSB) effects into the
theory of compound nuclear reactions.  We show that the only meaningful test
of TRSB in the overlapping resonances regime is through the study of
cross-section correlations.  The effect is channel-dependent.  In the isolated
resonance regime, we employ $K$-matrix theory to show the impact of TRSB
using the fact that when only one eigen-channel
participates in populating and depopulating the compound
resonance.
\end{abstract}
\vfill
\pacs{25.40.NY, 11.30.Er, 24.10.Ht, 24.80.Dc}

\newpage

\section{Introduction}

The breaking of CP symmetry observed for neutral kaons \cite{C} implies,
because of CTP invariance, time reversal symmetry breaking (TRSB), which has
never been demonstrated experimentally.  Because of the results of tests
looking for TRSB effects, expectations are that they will be small.  We
believe, however, that the amplification which occurs in low energy neutron
reactions may make them visible --- and in any event will provide
limits.\cite{HJ} \cite{B}.
This feasibility is suggested by the
recent observation of the violation of parity conservation using neutron
resonances \cite{F}.
In this note we present a summary of the reaction theory
needed to extract time symmetry breaking from both resonance and
energy averaged experiments.
In the following,
we will discuss first, overlapping resonances and, in the next section,
the isolated resonance regime.

\section{Overlapping Resonances}

We make use of the methods employed by Kawai, Kerman and McVoy (KKM)
\cite{KKM}
to obtain the average fluctuation cross-section.
Because of time symmetry breaking the $S$ matrix is not symmetric, $S_{ab}\neq
S_{ba}$.  In the case of neutron resonances, $S$ has a form
generalized from KKM, i.e.
\begin{equation}
 S_{ab}=\bar S_{ab} - i\sum_q \frac{g_{qa}\check g_{qb}}{E-E_q}
 =\bar S_{ab}+S_{ab}^{fl}
 \label{1}\end{equation}
In this equation $\bar S$ is the optical model $S$ matrix while the $g$'s are
given by
\begin{eqnarray}
 g_{qa} &=& \sqrt{2\pi}\langle\tilde q|H_{QP}|\psi^{(+)}_a\rangle\equiv
 g_{qa}^0+\Delta g_{qa}
 \label{2} \\
 \check g_{qa} &=&  \sqrt{2\pi}\langle\psi^{(-)}_a|H_{PQ}|q\rangle\equiv
 g_{qa}^0-\Delta g_{qa}
 \label{3}
\end{eqnarray}
where $\Delta g$ represents the effect of the breaking of time reversal
symmetry on the fluctuations.   Under time reversal invariance
$\Delta g = 0$.
Notice that both $g^0\/$  and $\Delta g\/$ are complex.

The complex energies
$E_q$, and the vectors $| q >\/$ and $< \tilde{q} |\/$
are solutions of the Schr\"odinger equations:
\begin{equation}
 \left(E-H_{QQ}-H_{QP}\frac{1}{E^+ -H_{PP}}H_{PQ}\right)|q\rangle=E_q|q\rangle
 \label{4}\end{equation}
and
\begin{equation}
 \left(E-H_{QQ}^+ -\left(H_{QP}\frac{1}{E^+ -H_{PP}}H_{PQ}\right)^+\right)
 |\tilde q\rangle=E_q^* |\tilde q\rangle
 \label{5}\end{equation}
Our notation is that of Ch.~III of the ``Theory of Nuclear
Reactions'' \cite{HF}.  The symmetry breaking
interaction is present in $H_{QQ}$, $H_{QP}$, $H_{PQ}$ and $H_{PP}$ since
\[
 H^T\neq H
 \]
where $H^T$ is the time reverse of $H$.  As one can see from Eq.~(\ref{2}) and
Eq.~(\ref{3}), symmetry breaking effects in nuclear reactions will have their
source in $H_{QP}$, $H_{PQ}$ and in $|q\rangle$.  In particular we note
$< r | q > \neq < \tilde{q} | r >^*\/$.
The resonance energies $E_q$,
even though they have been very slightly changed by TRSB,
will be averaged and will not affect the final result.

Assuming that $g^0$ and
$\Delta g$ are independently random.  The energy averaged
fluctuation cross-section is given by
\begin{equation}
 \langle\sigma_{ab}^{fl}\rangle=\langle S_{ab}^{fl} S_{ab}^{fl*}\rangle\ .
 \label{6}
\end{equation}
which, with the usual random phase assumption, gives
\begin{equation}
\bigl\langle \sigma^{f l}_{a b} \bigr\rangle =
           x^2_0 \bigl\langle g_{q a} \check{g}_{q b} g^*_{q a}
                      \check{g}^*_{q b} \bigr\rangle_q
\label{6a}
\end{equation}
We now introduce the following quantities obtained by applying the KKM
analysis \cite{KM}:
\begin{equation}
 X_{ab}=x_0\langle g_{qa}^0 g_{qb}^{0*}\rangle_q\ ,\ \
 x_{ab}=x_0\langle\Delta g_{qa}\Delta g_{qb}^*\rangle_q
 \label{7}
\end{equation}
In Eq.~(\ref{7})
$x_0 = \sqrt \frac{2 \pi}{\Gamma D}$, where $\Gamma\/$ and $D\/$are the
average width and spacing of the resonance.  This constant
will drop out in our final expressions.  The
average on the RHS of Eqs.~(\ref{6a}) and (\ref{7}) is
carried out over the compound nuclear states as indicated by the
subscript $q$.  In terms of these quantities, the energy averaged fluctuation
cross-section is given by
\begin{equation}
 \bigl\langle\sigma_{ab}^{fl}\bigr\rangle
           =(X+x)_{aa}(X+x)_{bb}+(X-x)_{ab}^2
 \label{8}
\end{equation}
where pair correlations among the $g_{ga}\/$'s
and the random independence of $g^0\/$ and $\Delta g\/$
have been assumed in performing the average in
Eq.~(\ref{6a}).  It is assumed that the number of levels is large and the
number of exit channels is large.

As in the KKM example the above quantities can be related to the transmission
coefficients which are given by the optical model
\begin{equation}
 T_{ab}\equiv
 \delta_{ab}-(\bar S\bar S^+)_{ab}=
           \Bigl\langle \bigl(S^{fl}S^{fl+} \bigr)_{ab} \Bigr\rangle
 \label{9}\end{equation}
where $\bar S$ is the optical model $S$ matrix.  The last expression can be
related to $X$ and $x$:
\begin{equation}
 T_{ab}=(X+x)_{ab}Tr(X+x)+\left( (X-x)^2\right)_{ab}
 \label{10}
\end{equation}

We can now formulate the consequences of the above analysis.  First note that
according to Eq.~(\ref{8}),
$\langle\sigma_{ab}^{fl}\rangle=\langle\sigma_{ba}^{fl}\rangle$.  An
early experimental test of this was reported in \cite{B}.
Thus it is not possible
to detect time reversal symmetry breaking by comparing the energy averaged
cross
sections for $a\to b$ with that for $b\to a$.  Detailed balance holds in the
presence of symmetry breaking.  To observe symmetry breaking
we have to analyze the appropriate cross-section correlation function
$C_{ab}$
\begin{equation}
 C_{ab}  = \frac{\bigl\langle\sigma_{ab}^{fl}\sigma_{ba}^{fl} \bigr\rangle
                 - \bigl\langle\sigma_{ab}^{fl} \sigma^{fl}_{ab} \bigr\rangle}
                 {\bigl\langle\sigma_{ab}^{fl}\bigr\rangle
                      \bigl\langle\sigma^{fl}_{ab}\bigr\rangle}
 \label{11}
\end{equation}
It can be shown using the pair correlation assumption that to first order in
$N^{-1}$, where $N$ is the number of open channels,
\begin{equation}
 C_{ab}= \frac{\bigl\langle\tilde\sigma_{ab}^{fl}\bigr\rangle^2
           -  \bigl\langle \sigma^{fl}_{ab} \bigr\rangle^2}
            { \bigl\langle\sigma_{ba}^{fl} \bigr\rangle^2 }
 \label{12}
\end{equation}
where we have introduced the pseudo-fluctuation cross-section
\begin{eqnarray}
  \bigl\langle\tilde\sigma^{fl}_{ab}\bigr\rangle
           &=& \bigl\langle S_{ab}S_{ba}^*\bigr\rangle
            \nonumber \\
 &=& (X-x)_{aa} (X-x)_{bb}+(X+x)_{ab}^2
 \label{13}
\end{eqnarray}
{}From Eq.~(\ref{8}), Eq.~(\ref{13}) and Eq.~(\ref{10}), neglecting the
non-diagonal terms $(X_{ab}=0=x_{ab})$, one finds, to leading order in the
TRSB matrix element, the following form for $C_{ab}$:
\begin{equation}
 C_{ab} \simeq
           -4  \frac{ x_{aa} X_{bb} + X_{aa}  x_{bb} }
                       {X_{aa} X_{bb} }  \>  ,
 \label{14}
\end{equation}
when written in terms of the transmission coefficient, we find
\begin{equation}
 C_{ab} \approx -4  \left(\frac{t_{aa}}{{\mathop{T}\limits^\circ}_{aa}}+
 \frac{t_{bb}}{{\mathop{T}\limits^\circ}_{bb}}\right)\ ,
 \label{15}
\end{equation}
where ${\mathop{T}\limits^\circ}_{aa}$ is the optical transmission matrix
element in channel $a$ without TRSB, and $t_{aa}$ is just the difference
$\bigl( T_{aa}-{\mathop{T}\limits^\circ}_{aa} \bigr)$.
We should mention here that the variance defined by
$V_{ab} = \bigl\langle \sigma^{fl}_{ab} \bigr\rangle^2 -
\bigl\langle \sigma^{fl}_{ab} \bigr\rangle^2\/$
has been carefully investigated by Kerman and Sevgen \cite{KS}.
These authors pointed out that $V_{ab}\/$ depends explicitly on $\Gamma / D\/$.
Therefore, to study TRSB, one should avoid using $V_{ab}\/$ and instead use
$C_{ab}\/$ where the $\Gamma/D\/$ factors cancel out.

Equation (\ref{15}) clearly shows that $C_{ab}$ depends on the channels.  This
is in contrast to the result of Refs.~\cite{E}, \cite{BB} and
\cite{BHW}.  In particular Ref.~\cite{BB} calculated
$|R_{ab}|^2 \equiv
           \frac{ \Bigl[ \bigl< \sigma^{fl}_{ab} \sigma^{fl}_{ba} \bigr>
                      - \bigl< \sigma^{fl}_{ab}  \bigr>^2 \Bigr]  }
                      { \bigl< \sigma^{fl}_{ab} \bigr>^2} \/$
and claimed that it
does not depend on $a\/$ and $b\/$.
The reason for this
is that the authors of \cite{E,BB,BHW}
consider TRSB to be entirely in $H_{QQ}$ and do
not consider its effect on $H_{PQ}$.  Thus, if
treated as purely internal mixing, the TRSB is channel independent,
both in $R_{ab}\/$ and $C_{ab}\/$.

The use of symmetry breaking one-body potentials to treat energy-averaged
observables has already been proven successful in the case of parity
non-conservation \cite{HC}, \cite{HKL1}.  An optical model description of TRSB
\cite{HG}, following the lines of \cite{HC},
has been presented recently.  It would be profitable to
calculate $C_{ab}$, Eq.~(\ref{15}), using such one-body models of TRSB.

\section{Isolated Resonances}

We turn now to the case of isolated resonances.  This situation is usually
encountered at neutron energies in the electron volt region.  The parity
non-conservation experiment of Ref.~\cite{F} was performed under these
conditions.  The study of TRSB in the isolated resonance regime has been
discussed recently \cite{VEB,VEB2}.  Here we present a different point of view
concerning this matter.
It is convenient for the discussion to use the $K$-Matrix, which
at a given isolated resonance, we write, in the presence of TRSB, as
\begin{equation}
 K_{cc'}^q =
           \frac{1}{2 \pi} \,
           \frac{\gamma_{qc}\gamma_{qc'}^*}{E-\epsilon_q}\> ,
 \label{16}
\end{equation}
$K^q$ is hermitian but neither
not real, nor is it symmetric.  Note the $\epsilon_q$ in (\ref{16})
is the real energy of the compound level, $q$.  The $T$-matrix is obtained
from the $K$-matrix through
\begin{equation}
 T^q_{cc'} =
           \frac{1}{2 \pi} \,
           \frac{\gamma_{qc} \gamma_{qc'}^*}{E-\epsilon_q+i\Gamma_q/2}\> ,
\qquad \hbox{with }
\Gamma_q =
           2 \pi \sum_c | \gamma_{q c} |^2 + \Gamma^\gamma_q \> ,
 \label{18}
\end{equation}
with $\Gamma^\gamma_q\/$ being the radiative decay width of resonance $q\/$,
which is the dominant piece of $\Gamma_q\/$.
The above form of $T^q_{cc'}\/$ establishes the link with the discussion at
the beginning
of the paper, {\it i.e.\/}~for the present case
$g^0_{q c} = R e \gamma_{q c} = \overcirc{\gamma}\/$ and
$\Delta g_{q c} = i I m \gamma_{q c} \equiv i \gamma^w_{q c}\/$.
Going back to Eq.~(\ref{16}), we introduce the eigenchannels that diagonalize
$K_{cc'}^q$, by the requirement
$\gamma_{qc} \sum\limits_{c'} \gamma_{qc'}^* f_{c'}=\lambda f_c$,
which is solved by $\lambda=\sum\limits_{c'} \left|\gamma_{qc'}\right|^2$ and
$f_c=\gamma_{qc}$.  All other solutions have $\lambda=0$.  Thus there is only
one physical eigenchannel for each level $q$.  We thus write for $K^q$ in
operator form
\begin{equation}
 K^q=\frac{\left|\gamma_q\right|^2}{E-\epsilon_q}|\hat\gamma_q\rangle
 \langle\hat\gamma_q|
 \label{19}\end{equation}
where $|\hat\gamma\rangle$ is a unit vector with components
$\gamma_{q c} / \sqrt{\sum\limits_c | \gamma_{q c} |^2}\/$.
Note that this eigenchannel also
diagonalizes $T^q$.
\begin{equation}
 T^q = \frac{|\gamma_q|^2}{E-\epsilon_q+i\Gamma_q/2}
 |\hat \gamma_q\rangle\langle\hat{\gamma}_q|\ .
 \label{20}\end{equation}
If we represent the TRSB measurement operator by $\theta_T$, then the
difference in total
cross-sections with two different neutron helicities and a polarized nuclear
target is
\begin{equation}
 \Delta\sigma_q=\frac{2\pi}{k^2}Im
 \frac{1}{E-\epsilon_q+i\frac{\Gamma_q}{2}}
 \sum_{c',c} \langle \gamma_{q c} |\theta_T| \gamma_{q c'} \rangle
 \label{22}
\end{equation}
where $c$ denotes the entrance channel, and $c'$ the channel that
$\theta_T$ couples.  Since $\theta_T$ is by definition Hermitian and
antisymmetric, $\langle \gamma_{qc}|\theta_T| \gamma_{qc'}\rangle$
must be purely imaginary.
Accordingly, we have at the $q$-th resonance,
\begin{equation}
 \Delta\sigma_q = \frac{2\pi}{k^2} \frac{2}{\Gamma_q}
\sum_{c, c'} \left[\gamma_{qc'}^* \gamma_{qc} -
           \gamma_{qc'} \gamma_{qc}^*\right]
(\theta_T)_{cc'} \> .
 \label{23}
\end{equation}
As noted above we have $\gamma_{q c} =
           \overcirc{\gamma}_{q c} + i \gamma_{q c}^W$,
where $\overcirc{\gamma}_{q c}$ is the (real) strong $T$-even amplitude.
Then to first order in $\gamma^W$, and defining
$(\theta_T)_{cc'} = i \theta_{cc'}\/$
where $\theta_{cc'}\/$ is antisymmetric
\begin{equation}
 \Delta\sigma_q = \frac{2\pi}{k^2} \, \frac{4}{\Gamma_q}
 \sum_{c c'} \left[\gamma_{q c}^W \overcirc{\gamma}_{q c'} \right]
 \theta_{c c'}
 \label{24}
\end{equation}
For the special case of two channels, one $(C_1)\/$ coupled weakly and the
other
$(C_0)\/$ coupled strongly
\begin{equation}
\Delta \sigma_q = \frac{2 \pi}{k^2} \frac{4}{\Gamma_q}
           \gamma^{W}_{qc_1} \overcirc{\gamma}_{qc_0}
\label{25}
\end{equation}
The asymmetry,  $P_q \equiv \frac{\Delta\sigma_q}{2\sigma_q}$ is then given by
(ignoring the background contribution)
\begin{equation}
P_q = \gamma^W_{qc_1} / \overcirc{\gamma}_{qc_0}
\label{26}
\end{equation}

Generally, $P_q\/$ will have a vanishing average value because of the random
nature of $\gamma^W_{qc}\/$, and $\overcirc\gamma_{qc_0}\/$.  This situation
changes if a local $2p - 1h\/$ doorway dominates the TRSB mixing.  Just
as in the PNC case
\cite{F}, whose fine structure has been recently analyzed in \cite{HKL2},
the $P_q\/$,
Eq.~(\ref{26}), will have a definite sign.

We mention here that the detailed nature of the $T$-violation experiment
depends on the $T$-violating operator $\theta_T$.  Several forms may be cited.
For parity non-conserving, time reversal violating, these are
\begin{eqnarray}
            \theta_{TP,1} &=&  \left( \vec\sigma \cdot \hat q \right)
 \label{28} \\
            \theta_{TP,2} &=&  \left( \vec\sigma \times \vec I \right) \cdot
            \hat p \> .
 \label{29}
\end{eqnarray}
The time reversal violating $P\/$-even interactions are more complicated.
 We list
these operators in terms of the unit vectors
$\hat{q} = \hat{k'} - \hat{k}\/$, $\hat{p} = \hat{k} + \hat{k'}\/$ and
$\hat{n} = \hat{p} \times \hat{q}\/$
\begin{eqnarray}
 \theta_{T,1} &=& i
           [\vec\sigma \times \vec{ I} \cdot \hat{n}]
 \label{30} \\
 \theta_{T,2} &=&
           (\vec\sigma \times \vec{ I}  \cdot \hat q)
           (\vec I \cdot \hat q)
            \label{31} \\
 \theta_{T,3} &=&
           \left (\vec\sigma \times \vec I \cdot \hat p)
           (\vec I \cdot \hat p \right )
           \label{32}
\end{eqnarray}
where $\vec\sigma/2$ is the spin of the nucleon, $\vec I$ the spin of the
target nucleus.  In a neutron
transmission experiment where the total cross sections are measured, only
$\theta_{TP,2}$ and $\theta_{T,3}$ survive.  It is clear that in order to see
time reversal violation both for $P$-odd and $P$-even, we must have at least
two channel spins coupled by the violating interaction.  As examples we
mention the transition $^1P_1\to {^3S_1}$ caused by the $T$-odd, $P$-odd
interaction and the transition ${^1P_1} \to {^3P_1}$
caused by a $P$-even, $T$-odd
interaction.  In particular, the transition $^1P_1\to {^3S_1}$ mentioned above,
which could occur in the neutron scattering from a spin $1/2$ nucleus, is
particularly interesting as it resembles the $P$-odd $T$-even case studied in
Ref.~\cite{F} except for the change in channel spin.  We suggest that to get a
measurable $P_q$, Eq.~(\ref{28}), one comes in a $^1P_1$ state in a nucleus
where the single particle $P$-wave
strength function exhibits a minimum and comes
out in a centrifugal barrier uninhibited
$^3S_1$-state sitting at a maximum in the
corresponding $s$-wave
strength function.  In the $A\simeq 180$ and 140 region one
encounters such a situation \cite{SKM}.  The nucleus $^{139}La$ considered in
Ref.~\cite{TSMY} seems to be a good candidate to study TRSB.

\section{Conclusions}

In conclusion, we have discussed the theory of TRSB in nuclear reactions.  We
first analyzed the case of overlapping resonances and showed that the
cross-section correlation function $C_{ab}^2$ depends explicitly on the
channels, contrary to what has been suggested previously.  We then discussed
the low-energy (eV's) isolated resonances case and derived a $P_q\/$
expression for
the asymmetry $P_q\/$ at a given resonance.  A more detailed
account of the present work will appear elsewhere \cite{FHK}.


\begin{references}

\bibitem{C}
C.R.~Christensen {\it et~al.\/}, {\it Phys.~Rev.~Lett.\/} {\bf 13} 138
(1964).

\bibitem{HJ}
E.M.~Henley and B.A.~Jacobson, {\it Phys.~Rev.\/} {\bf 113} 225 (1959).

\bibitem{B}
For a recent review, see:
W.C.~Haxton and A. H\"oring, {\it Nucl.~Phys.\/} {\bf A560} 469 (1993);\\
P.~Herczeg, in {\it Tests of Time Reversal Invariance in Neutron Physics},
ed.~N.R.~Roberson, C.R.~Gould and J.D.~Bowman (World Scientific, Singapore,
1987), p.~24.

\bibitem{F}
C.M.~Frankle {\it et~al.\/}, {\it Phys.~Rev.~Lett.\/} {\bf 67} 564 (1991).

\bibitem{KKM}
M.~Kawai, A.K.~Kerman and K.W.~McVoy, {\it Ann.~Phys.\/} (NY) {\bf 75} 156
(1973).

\bibitem{HF}
H.~Feshbach, {\it Theoretical Nuclear Physics:  Nuclear Reactions\/} (John
Wiley and Sons, New York, 1992).

\bibitem{KM}
A.K. Kerman and K.W. McVoy, {\it Ann. Phys.\/} (NY) {\bf 122} 197 (1979).

\bibitem{KS}
A.K. Kerman and A. Sergen, {\it Ann. Phys.\/} (NY) {\bf 102} 570 (1976).

\bibitem{E}
T.E.O.~Ericson, {\it Phys.~Lett.\/} {\bf 23} 97 (1966).

\bibitem{BB}
E.~Blanke, H.~Driller, W.~Gl\"ockle, H.~Genz, A.~Richter and G.~Schreider,
{\it Phys.~Rev.~Lett.\/} {\bf 51} 355 (1983).

\bibitem{BHW}
D.~Boos\'e, H.L.~Harney and H.A.~Weidenm\"uller, {\it Phys.~Rev.~Lett.\/} {\bf
19} 2012 (1986);\\
E.D.~Davis and U.~Hartmann, {\it Ann.~Phys.\/} (NY) 211 (1991) 334.

\bibitem{HC}
B.V.~Carlson and M.S.~Hussein, {\it Phys.~Rev.\/} {\bf C47} 336 (1993).

\bibitem{HKL1}
M.S.~Hussein, A.K.~Kerman and C.-Y.~Lin, submitted for publication.

\bibitem{HG}
V.~Hnizdo and C.R.~Gould, {\it Phys.~Rev.\/} {\bf C49} R612 (1994).

\bibitem{VEB}
V.E.~Bunakov, {\it Phys.~Rev.~Lett.\/} {\bf 60} 2250 (1988).

\bibitem{VEB2}
V.E.~Bunakov, E.D.~Davis and H.A.~Weidenm\"uller, {\it Phys.~Rev.\/} {\bf C42}
1718 (1990).

\bibitem{HKL2}
M.S.~Hussein, A.K.~Kerman and C.-Y.~Lin, {\it Z. Phys.~A\/} in press.

\bibitem{M}
F.C.~Michel, {\it Phys.~Rev.\/} {\bf B329} 133 (1964).

\bibitem{SKM}
S.K.~Mughabghab, {\it Neutron Cross Sections\/} (Academic, Orlando, 1984)
Vol.~1.

\bibitem{TSMY}
Y.~Takahashi, H.M.~Shimizu, T.~Maekawa and T.~Yabuzaki, {\it Phys.~Lett.\/}
{\bf B326} 27 (1994).

\bibitem{FHK}
H.~Feshbach, M.S.~Hussein and A.K.~Kerman, in preparation.

\end{references}
\end{document}